\documentclass[preprint2,12pt]{proto}
\usepackage{times}
\usepackage{natbib}

\begin{document}

\title{\textbf{\LARGE Asteroid Systems: Binaries, Triples, and Pairs}}

\author {\textbf{\large Jean-Luc Margot}}
\affil{\small\em University of California, Los Angeles}
\author {\textbf{\large Petr Pravec}}
\affil{\small\em Astronomical Institute of the Czech Republic Academy of Sciences}
\author {\textbf{\large Patrick Taylor}}
\affil{\small\em Arecibo Observatory}
\author {\textbf{\large Beno\^it Carry}}
\affil{\small\em Institut de M\'ecanique C\'eleste et de Calcul des \'Eph\'em\'erides}
\author {\textbf{\large Seth Jacobson}}
\affil{\small\em C\^ote d'Azur Observatory}

\begin{abstract} %
\begin{list}{ } {\rightmargin 0.5in}
\baselineskip = 11pt
\parindent=1pc {\small In the past decade, the number of known binary
  near-Earth asteroids has more than quadrupled and the number of
  known large main belt asteroids with satellites has doubled.  Half a
  dozen triple asteroids have been discovered, and the previously
  unrecognized populations of asteroid pairs and small main belt
  binaries have been identified.  The current observational evidence
  confirms that small ($\lesssim$20 km) binaries form by rotational
  fission and establishes that the YORP effect powers the spin-up
  process.  A unifying paradigm based on rotational fission and
  post-fission dynamics can explain the formation of small binaries,
  triples, and pairs.  Large ($\gtrsim$20 km) binaries with small
  satellites are most likely created during large collisions.
\\~\\~\\~}%

\end{list}
\end{abstract}  %

\maketitle

\centerline{\textbf{ 1. INTRODUCTION}}
\bigskip
\noindent
{\textbf{ 1.1. Motivation}}
\bigskip

Multiple-asteroid systems are important because they represent a
sizable fraction of the asteroid population and because they enable
investigations of a number of properties and processes that are often
difficult to probe by other means.  
The binaries, triples, and pairs
inform us about a great variety of asteroid attributes, including
physical 
properties, composition, interior
structure, formation processes, and evolutionary processes.

Observations of binaries and triples provide the most powerful way of
deriving reliable masses and densities for a large number of objects.
The density measurements help us understand the composition and
internal structure of minor planets.
Binary systems offer opportunities to measure
thermal
and mechanical
properties, which are generally poorly known.

The binary and triple systems within near-Earth asteroids (NEAs), main
belt asteroids (MBAs), and trans-Neptunian objects (TNOs) exhibit a
variety of formation mechanisms
~\citep{merl02,noll08}.  As such, they provide an invaluable window on
accretional, collisional, tidal, and radiative processes that are
critical in planet formation.  The distribution and configurations of
the multiple-asteroid systems also provide a rich array of constraints
on their environment, their formation, and their evolutionary
pathways.

Observations rely primarily on ground-based telescopes and the Hubble
Space Telescope (HST).  For an up-to-date list of binaries and triples
in the solar system, see \citet{john14}.  We describe observational
techniques only briefly because this material is available
elsewhere~\citep[e.g.,][]{merl02}.  A few emerging techniques will be
described in more detail.  Likewise, we refer the reader to other
texts for an extensive history of the field~\citep[e.g.,][]{merl02}
and highlight only a few of the developments here.

\bigskip
\noindent
{\textbf{ 1.2. History}}
\bigskip

Early search programs for asteroid satellites were unsuccessful,
returning negative or dubious results, such that the authors of the
{\em Asteroids II} review chapter chose the prudent title ``Do
asteroids have satellites?'' \citep{weid89}.  The chapter provides an
excellent discussion of the physics of several formation mechanisms
that were postulated at the time.  The perspective changed with the
flyby of (243) Ida by the Galileo spacecraft in 1993 and the discovery
of its small satellite Dactyl~\citep{chap95,belt95}.  
Ground-based efforts intensified and resulted in the discovery of a
satellite around (45) Eugenia by~\citet{merl99}.  Several other
discoveries followed in rapid succession.  The relatively small sizes
of the MBA satellites suggested formation in sub-catastrophic or
catastrophic collisions~\citep{durd96,dore97}.

The discovery of MBA satellites, coupled with
analysis of terrestrial doublet craters~\citep{bott96,bott96i} and
anomalous lightcurve observations~\citep{prav97}, suggested the
existence of binary asteroids in the near-Earth population as well.
The unambiguous detection of five NEA binaries by radar cemented this
finding and indicated that NEA satellites form by a spin-up and
rotational fission process~\citep{marg02s}.  Lightcurve observers
reached the same conclusion independently~\citep{prav07}.
Both radar and lightcurve observations revealed that, far from being
rare, binary asteroids are relatively
common~\citep{prav99,marg02s,prav06}.  By the time the {\em Asteroids
  III} review chapter was written, a more decisive title (``Asteroids
do have satellites'') had become appropriate \citep{merl02}.  This
review focuses on the developments that followed the publication of
{\em Asteroids III}.

\bigskip
\noindent
{\textbf{ 1.3. Terminology}}
\bigskip

Two- and three-component asteroids that are gravitationally bound will
be referred to as {\em binary asteroids} (or {\em binaries}) and {\em
  triple asteroids} (or {\em triples}), respectively.
({\em Triple} is favored over the more directly analogous terms {\em
  trinary} and {\em ternary} because of long-established usage in
astronomy).  {\em Asteroid pairs} denote asteroid components that are
genetically related but not gravitationally bound.  {\em Paired
  binaries} or {\em paired triples} are asteroid pairs where the
larger asteroid is itself a binary or triple asteroid.  The larger
component in binaries, triples, and pairs is referred to as the {\em
  primary component} or {\em primary}.  The smaller component in
binaries is referred to as the {\em secondary component} or {\em
  secondary}.

There has been some confusion in the literature about the meaning of
the word ``asynchronous.''  
Here, we adopt the terminology proposed by \citet{marg10poland} and
later implemented by \citet{jaco11icarus} and \citet{fang12spinorbit}.
Binaries with an absence of spin-orbit
synchronism are called {\em asynchronous binaries}.  Binaries with a
secondary spin period synchronized to the mutual orbit period are
called {\em synchronous binaries}.  Binaries with both primary and
secondary spin periods synchronized to the mutual orbit period are
called {\em doubly synchronous binaries}.
If generalization to systems with more than one satellite is needed,
we affix the terms {\em synchronous} and {\em asynchronous} to the
satellites being considered.

It is useful to present results for {\em small} and {\em large}
asteroids.  We place an approximate dividing line at the size at which
objects are substantially affected by the YORP effect during their
lifetime.
For typical NEAs and MBAs, this dividing line corresponds to a
diameter of about 20~km~\citep{jaco14dist}.  We define {\em very
small} asteroids as those with diameters of less than 200 m.  This is
the approximate size below which many asteroids are observed to spin
faster than the disruption rate of a body with no shear or tensile
strength $\omega_d = \sqrt{4 \pi \rho G / 3}$, where $\rho$ is the
density and $G$ is the gravitational constant.

We use two additional acronyms.  The YORP effect is a
radiation-powered rotational acceleration mechanism for small
asteroids~\citep{Rubincam:2000fg}.  The binary YORP (BYORP) effect is
a radiation-powered acceleration mechanism that may expand or contract
the orbits of some synchronous asteroids~\citep{cuk05}.

\bigskip

\centerline{\textbf{ 2. OBSERVATIONS}}
\bigskip

Several observational techniques are available for discovering,
detecting, and studying binaries, triples, and pairs, each with its
strengths and weaknesses.
This section describes recent
results and illustrates the complementarity of the observational
techniques that characterize individual asteroid systems and entire
populations.  

\bigskip
\noindent
\textbf{ 2.1 Radar Observations of NEA Systems}  %
\bigskip

Radar has proven 
to be a powerful method of detecting secondaries to NEAs, 
enabling the discovery (as of September 2014) of the satellites in
71\% of the 49 known multiple-component NEA systems, including 33 of
47 binaries and both undisputed triple systems. Of the 14 binary NEAs
discovered via optical lightcurve techniques, 6 have been confirmed
with follow-up radar observations during later apparitions.
Overall, radar detections suggest that about one in six NEAs larger
than 200 m in diameter are multiple-asteroid
systems~\citep{marg02s,tayl12dps}, though 200~m is not a sharp
cutoff. Three binary NEA systems identified by radar have primary
components with suggested diameters of 120 m to 180 m: 2003
SS$_{84}$~\citep{nola03}, (363599) 2004 FG$_{11}$~\citep{tayl12fg11},
and 1994 CJ$_{1}$~\citep{tayl14dps}. For comparison, the largest
primaries of binary NEAs imaged with radar: (5143)
Heracles~\citep{tayl12heracles}, the possible triple (276049) 2002
CE$_{26}$~\citep{shep06}, and (285263) 1998 QE$_{2}$~\citep{spri14},
are more than an order of magnitude larger at $>$3 km in diameter.  It
is likely that $\sim$8 km diameter (1866) Sisyphus has a secondary
based on analysis of frequency-only observations obtained on four
separate dates in 1985 (Ostro, pers.\ comm., 2001).

\begin{figure}
\includegraphics[angle=90,width=\columnwidth]{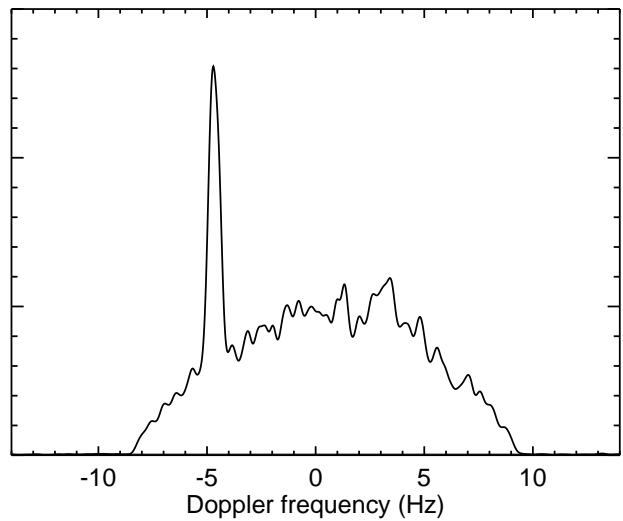}
\includegraphics[angle=0,width=\columnwidth]{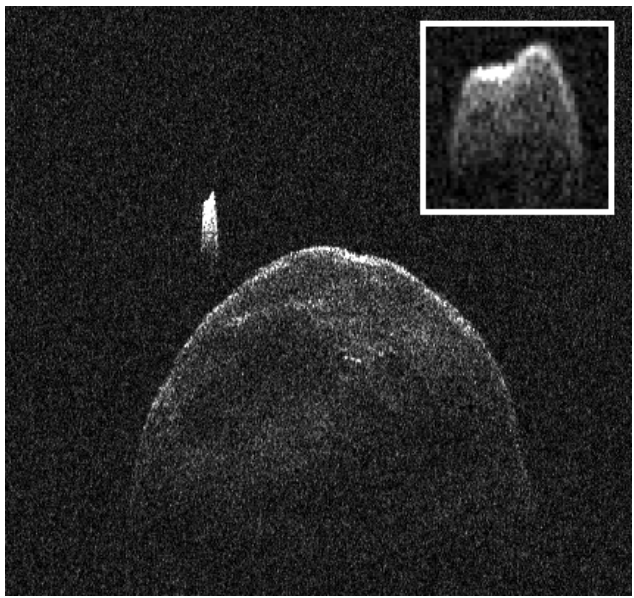}
\caption{\small Binary near-Earth asteroid (285263) 1998 QE$_{2}$ as
detected using the Arecibo planetary radar system. In the
frequency-only spectrum showing echo power as a function of Doppler
frequency (top), the narrowband echo of the tidally locked secondary
stands out against the broadband echo of the larger, faster-rotating
primary.
In the radar image (bottom), the components are spatially resolved
(7.5 m/pixel).  The vertical axis represents distance from the
observer increasing downward.
The horizontal axis is Doppler frequency due to the orbital and
rotational motion of the components.  Note that if one
summed the pixel values in each column of the image, the intensity as
a function of Doppler frequency would
approximate the spectrum above. The secondary is roughly one-fourth
the size of the primary (measured in the vertical dimension), though
the Doppler breadth of the primary gives the illusion of a greater
size disparity. The shape of the secondary (inset) is distinctly
nonspherical when viewed with finer frequency resolution.
\label{fig-radar}}
\end{figure}

Radar observations can be used to detect asteroid satellites because
of the ability to resolve the components of the system both spatially
(along the observer's line of sight) and in terms of frequency (due to
Doppler shifts from the rotational and orbital line-of-sight
velocities), resulting in a measurable separation between the
components in two dimensions.
Direct detection of a satellite in frequency-only spectra or radar
images typically occurs within one observing session and often within
minutes of observation. The bandwidth of the echo of a component
scales directly with the diameter and rotation rate. Thus, in a
frequency-only experiment, the signal of the smaller, relatively
slowly rotating satellite is condensed to a smaller bandwidth that is
superimposed upon the broadband signal of the larger, 
often rapidly rotating, primary (Fig.~\ref{fig-radar}, top).
Not all radar-observed binaries present this characteristic spectrum
(e.g., where the secondary spins faster than the synchronous rate), but
all are readily detected in radar images when the components are also
resolved spatially (Fig.~\ref{fig-radar}, bottom).
Because the spatial resolution achieved with radar instruments
corresponds to an effective angular resolution of better than $\sim$1
milliarcsecond (mas), there is no bias against the detection of
satellites orbiting very close to the primary component.  Multiple
measurements of the range and frequency separations of the components
over days of sky motion provide the geometric leverage required to
determine the orbit of the secondary around the primary. 
This can be done for any orbital orientation and yields the 
total system mass, a property that is difficult to estimate otherwise.
Other techniques involve analyzing spacecraft flyby and orbit
trajectories~\citep[e.g.,][]{yeom99},
measuring the Yarkovsky orbital drift in conjunction with thermal
properties~\citep[e.g.,][]{ches14}, or observing the gravitational
perturbations resulting from asteroid
encounters~\citep[e.g.,][]{hilt02}.

Most binary NEA systems observed to date have a rapidly rotating
primary and a smaller secondary of order a few tenths the size of the
primary (a secondary-to-primary mass ratio of roughly 0.001 to 0.1),
whose rotation is synchronized to the mutual orbit period. The
majority of primaries rotate in less than 2.8 h, though they range
from 2.2593~h for (65803) Didymos~\citep{prav06} to 4.749~h for 1998
QE$_{2}$ (P. Pravec, pers.\ comm., 2013).  The known outlier is the
nearly equal-mass binary (69230) Hermes, whose components both appear
to have 
13.894 h periods synchronized to their mutual orbit period
\citep{marg06iau}.  This doubly synchronous configuration is most
likely due to rapid tidal evolution~\citep{tayl11}. While the
rotations of satellites in NEA binaries tend to be tidally locked to
their orbital mean motions with periods typically within a factor of
two of 24 h (often resulting in the characteristic appearance shown
in Fig.~\ref{fig-radar}), about one in four radar-observed
multiple-asteroid systems have an asynchronous
satellite~\citep{broz11}, all of which rotate faster than their
orbital rate. 
Well-studied examples include (35107) 1991 VH~\citep{naid12dda},
(153958) 2002 AM$_{31}$~\citep{tayl13}, (311066) 2004
DC~\citep{tayl08}, and the outer satellites of both undisputed triple
systems (153591) 2001 SN$_{263}$~\citep{nola08dps,fang11,beck15} and
(136617) 1994 CC~\citep{broz11,fang11}.
Of the known asynchronous satellites, all have wide component
separations ($>$7 primary radii), translating to longer-than-typical
orbital periods, and/or eccentric orbits
($>$0.05), that are either remnants of their formation mechanism or
products of subsequent dynamical evolution~\citep{fang12spinorbit}.

\enlargethispage{0.5cm}

The shortest orbital periods detected with radar so far are those of
Didymos and 2006~GY$_2$ with $P_{\rm orb} = 11.90_{-0.02}^{+0.03}$~h
and $11.7 \pm 0.2$~h, respectively~\citep{benn10,broo06}.  For
Didymos, the semi-major axis is $a = 1.18_{-0.02}^{+0.04}$~km, just
outside the classical fluid Roche limit of $\sim$1 km for
equal-density components.  Other systems with satellites orbiting near
this limit include 2002 CE$_{26}$ and 2001 SN$_{263}$.  The
significance of this limit is unclear, as $\sim$100 m secondaries with
a cohesion comparable to comet regolith or sand can likely survive on
orbits interior to the Roche limit~\citep[][and references
  therein]{tayl10}.

Inversion of a series of radar images can provide a three-dimensional
shape model and complete spin-state description given sufficient
signal, resolution, and orientational coverage~\citep{huds93,magr07}.
Shape reconstruction of the larger component of (66391)
1999~KW$_{4}$~\citep{ostr06}
demonstrated that the canonical shape of an NEA primary has a
characteristic circular equatorial bulge, uniformly sloped sides, and
polar flattening akin to a spinning top. Such a shape is shared by the
primaries of 2004~DC, 1994 CC, 2001 SN$_{263}$, and (185851) 2000
DP$_{107}$~\citep{naid15dp}, though some primaries have less
pronounced equatorial belts, e.g., 2002 CE$_{26}$ and 1998 QE$_{2}$.
Some single asteroids have a similar shape, e.g., (101955)
Bennu~\citep{nola13} and (341843) 2008~EV$_{5}$~\citep{busc11}, but do
not have satellites, 
possibly because one has not yet formed or has been lost in the past.
Shape model renditions are shown in Benner et al.\ (this volume).
Often the resolution of radar images
of the smaller satellites is insufficient for shape inversion, but
radar images suggest that the satellites are typically elongated, e.g.,
2000 DP$_{107}$, 1999 KW$_{4}$, 2001 SN$_{263}$, 1991 VH, and 1998
QE$_{2}$.

Shapes and volumes obtained from 
inversion of radar images, combined with the system mass derived from
the orbital motion observed in radar images, provide the density of
the system (or of the individual components if the mass ratio is
measurable from reflex motion).
Low densities of order 1 g/cm$^{3}$~\citep{shep06,beck15} 
to 2 g/cm$^{3}$~\citep{ostr06,broz11} suggest
significant internal macroporosity of order 50\%, implying a
rubble-pile internal structure for the components. At such low
densities, the rapid rotation of the primary places particles along
the equatorial belt in a near-weightless environment. The combination
of rapid rotation, shape, and implied porosity and rubble-pile
structure has implications for the formation mechanism of small
multiple-asteroid systems (Section 4).

While radar allows for direct, unambiguous detection of asteroid
satellites, its range is limited. Because radar requires the
transmission and reception of a signal, the strength of the received
signal falls as the fourth power of the distance to the target and,
thus, is best suited for detecting multiple-component systems passing
within $\sim$0.2 astronomical units (au) of Earth. Satellites in the
main asteroid belt simply tend to be too small and too far away to
detect with present radar capabilities and require application of
different observational techniques.

\bigskip
\noindent
\textbf{ 2.2 Lightcurve Observations of NEA and Small MBA Systems}
\bigskip

\begin{figure*}[ht]
  \includegraphics[angle=0,width=\textwidth]{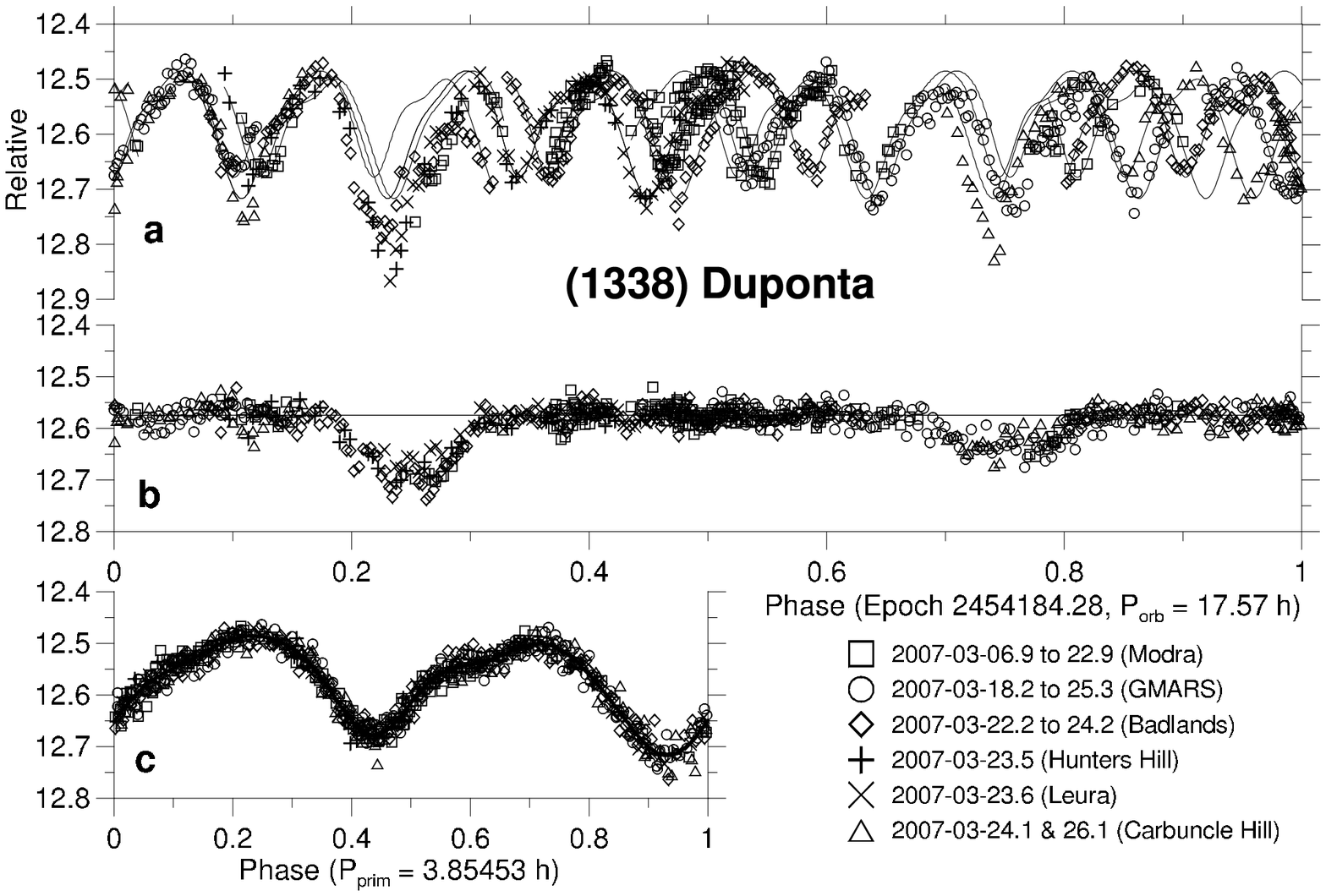}
  \caption{\small Lightcurve data of (1338) Duponta, which has a
    secondary-to-primary diameter ratio of about 0.24.  (a) The
    original data showing both lightcurve components, folded with the
    orbit period. (b) The orbital lightcurve component, derived after
    subtraction of the primary lightcurve component, showing the
    mutual events between components of the binary system.  (c) The
    primary lightcurve component.
    Figure from~\citet{prav12}.
    \label{fig-lc}}
\end{figure*}

A photometric lightcurve is a time series of measurements of the total
brightness of an asteroid.  Detections of binary asteroids by
photometric lightcurve observations utilize the fact that the
components can obscure or cast a shadow on one another, producing
occultations or eclipses, respectively.  The attenuations can be used
to both reveal and characterize binaries (Fig.~\ref{fig-lc}).  The
observational, analysis, and modeling techniques were described in
\citet{prav06, sche09lc, sche15}.

Early reports~\citep{tede79,cell85} of asteroids suspected to be
binaries on the basis of anomalous lightcurves (including (15)
Eunomia, (39) Laetitia, (43) Ariadne, (44) Nysa, (49) Pales, (61)
Danae, (63) Ausonia, (82) Alkmene, (171) Ophelia, and (192) Nausikaa)
have remained largely unconfirmed despite extensive follow-up
searches.  The first serious candidate for detection with this
technique was NEA (385186) 1994 AW$_{1}$~\citep{prav97}, whose binary
nature was confirmed by photometric observations in
2008~\citep{birl10}.  Since 1997, nearly 100 binaries among near-Earth
and small main belt asteroids
have been detected with the photometric method.  
The binary asteroid database constructed by~\citet{prav07} 
\\(http://www.asu.cas.cz/$\sim$asteroid/binastdata.htm) includes data
for 86 MBA and NEA binaries that were securely detected by photometry
and for which basic parameters have been derived, such as the primary
spin period, the orbital period, and the primary-to-secondary mean
diameter ratio.  A few tens of additional MBAs and NEAs are suspected
to be binaries and await confirmation with more detailed observations
in the future.

Among the main findings obtained from photometric observations is that
binary asteroids are ubiquitous.  They have been found among NEAs,
Mars-crossers (MCs), and throughout the main belt, both among
asteroids that have been identified as family members and among
asteroids that have not.
\citet{prav06} derived the fraction of binaries among NEAs larger than
300 meters to be $15 \pm 4$\%.  A binary fraction among MBAs has not
been derived precisely due to less well-characterized observational
selection effects, 
but their photometric discovery rate is similar to the discovery rate
of binaries among NEAs.  Thus, binaries are 
suspected to be 
as frequent among MBAs as they are among NEAs.  There appears to be an
upper limit on the primary diameter for photometrically detected
binaries of about 13 km; the largest 
detected binary is (939) Isberga with $D_p = 13.4 \pm 1.3$~km
\citep{carr15}.  A lower size limit on the primary
diameter $D_p$ is less clear.  The smallest
detected binary is 2000 UG$_{11}$ with $D_p = 0.26 \pm
0.03$~km~\citep{prav06},
but smaller binaries are known to exist (Section 2.1).  Their absence
in lightcurve data sets may be due in part to a bias against detecting
small binaries in the initial surveys.

\begin{figure}[!t]
  \includegraphics[angle=0,width=\columnwidth]{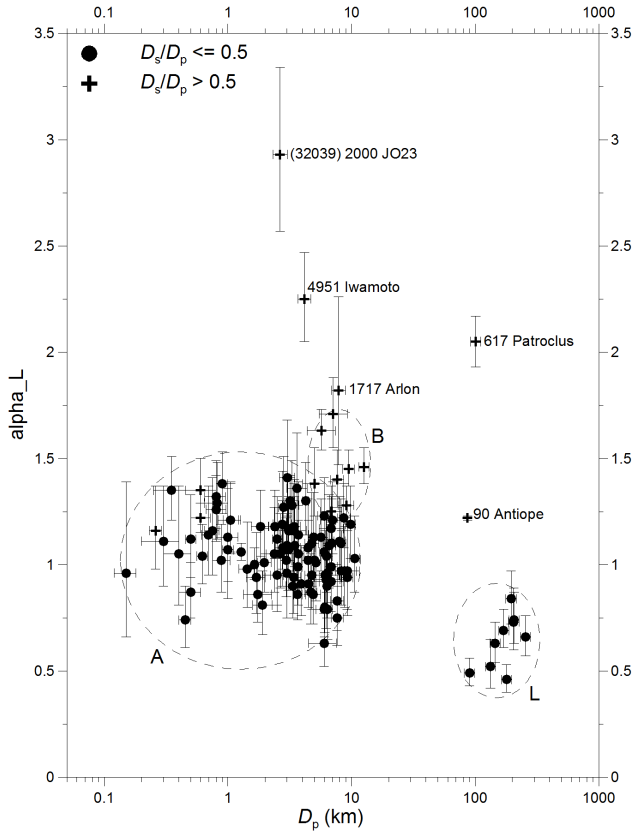}
  \caption{\small Estimated values of the normalized total angular
    momentum content of binaries 
    versus primary diameter.
    The quantity $\alpha_L$ is the sum of orbital and spin
    angular momenta normalized by the angular momentum of an
    equivalent sphere spinning at the critical disruption spin rate
    $\omega_d = \sqrt{4 \pi \rho G / 3}$ where $\rho$ is the density
    and $G$ is the gravitational constant.  In the Darwin notation,
    $\alpha_L = 1$ corresponds to $J/J' = 0.4$. Group A contains small
    NEA, MC, and MBA binaries.  Group B consists of doubly synchronous
    small MBAs with nearly equal-size components.  Group L represents
    large MBAs with small satellites (Section 2.5). Two exceptional
    cases are the doubly synchronous asteroids (90) Antiope and (617)
    Patroclus (Section 2.5).  Figure updated from~\citet{prav07}.
    \label{fig-angmom}}
\end{figure}

Another key finding is that small binary asteroids have, with only two
or three exceptions, a near-critical angular momentum content
(Fig.~\ref{fig-angmom}).  As shown by \citet{prav07}, their angular
momentum is consistent with
formation by fission of critically spinning parent bodies of a
cohesionless, rubble pile structure.  The exceptions are the semi-wide
systems (32039) 2000 JO$_{23}$ and (4951) Iwamoto, and possibly also
(1717) Arlon, with orbital periods of 117 h to 360 h
and super-critical total angular momentum content.

The orbital poles of main belt binaries were found to have a highly
anisotropic distribution, concentrating within 30 degrees of the poles
of the ecliptic~\citep{prav12}.
The preferential 
orientations of the orbital poles suggest that their parent bodies or
the primaries were tilted by the YORP effect towards the asymptotic
spin states near obliquities 0 and 180 degrees, consistent with
observations of single asteroids~\citep{hanu11}.

Another significant finding is that there appears to be a lower limit
on the separation between components of binary systems of about $a/D_p
= 1.5$, corresponding to an orbital period of 11--12~h for typical
densities.  Lightcurve observations indicate that the orbital period
of Didymos is $P_{\rm orb} = 11.91 \pm 0.02$~h \citep{prav06},
consistent with the radar estimate.  This suggests an orbit close to
the Roche limit for strengthless satellites (but see prior remark
about orbits interior to the Roche limit).

Photometric observations of a binary system over multiple apparitions
can be used to detect a change in the separation of the components due
to the effect on mutual event timing.
An extensive set of photometric observations of the synchronous binary
(175706) 1996 FG$_3$ obtained during 1996-2013 places an upper limit
on the drift of its semi-major axis that is one order of magnitude
less than estimated on the basis of the BYORP theory~\citep{sche15}.
This system may be in an equilibrium between BYORP and tidal torques as
proposed for synchronous binary asteroids by \citet{jaco11apj}.

Some data sets strongly suggest the presence of triple asteroids.  In
these cases, an additional 
rotational component that does not belong to the primary or the close
eclipsing secondary is present in the lightcurve.  This additional
rotational component does not disappear during mutual events where the
eclipsing close secondary is obscured by or in the shadow of the
primary.  \citet{prav12} identified three such cases: (1830) Pogson,
(2006) Polonskaya, and (2577) Litva.  The latter has been confirmed by
direct imaging observations of the third body (second satellite) on a
wide orbit~\citep{merl13}.

Other data sets reveal the existence of paired binaries/triples.
Two such cases have been published: the pair composed of (3749) Balam
and 2009 BR$_{60}$~\citep[][and references therein]{vokr09} and the
pair composed of (8306) Shoko and 2011 SR$_{158}$~\citep{iauc9268}.
Balam is a confirmed triple, with a distant satellite detected by
direct imaging \citep{iauc7827} and a close satellite detected by
lightcurve observations~\citep{iauc8928}.  Shoko is a suspected triple
as well: Using lightcurve observations, \citet{iauc9268} detected an
eclipsing, synchronous close satellite with $P_{\rm orb} = 36.2$~h and
a third
rotational component attributed to an outer satellite.

\begin{figure}[!t]
  \includegraphics[angle=0,width=1.0\columnwidth]{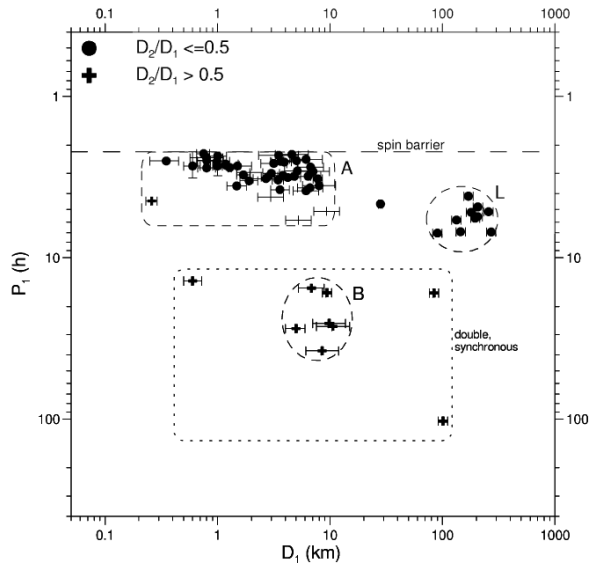}
  \caption{\small Primary rotation period versus primary
    diameter. Groups A, B, and L are defined in the caption of
    Fig.~\ref{fig-angmom}. Three doubly synchronous asteroids with
    nearly equal-size components lie isolated in the plot: (69230)
    Hermes on the left and (90) Antiope and (617) Patroclus on the
    right of group B.  Note that members of group A cluster near the
    disruption spin limit for strengthless bodies.  Figure 
    from~\citet{prav07}.
    \label{fig-abl}}
\end{figure}

While the population of binary NEAs and small MBAs is composed
primarily of synchronous systems, and secondarily of asynchronous
systems with low secondary-to-primary size ratios ($D_s/D_p < 0.5$),
doubly synchronous binaries with nearly equal-size components also
exist (Fig.~\ref{fig-abl}).  Nine such systems with $D_s/D_p > 0.7$
and orbital periods between 15 h and 118 h have been reliably
identified in the main belt~\citep[e.g.,][see also the Pravec and Harris binary database described above]{behr06,krys09}.

Another important observation is that, with the exception of doubly
synchronous systems, all binaries have unelongated, near-spheroidal
primary shapes, as evidenced by their low primary amplitudes not
exceeding 0.3~mag (when corrected to zero phase angle).  This suggests
that their primaries may have shapes similar to the top-like shapes
that have been observed for 1999 KW$_4$~\citep{ostr06} and
several other binaries by radar.

All the properties revealed by photometric observations indicate that
binary systems among NEAs and small MBAs were formed from critically
spinning cohesionless parent bodies, with YORP as the predominant
spin-up mechanism.  This finding is consistent with the fact that the
observed 0.2--13 km size range of binaries corresponds to the size
range where the spin barrier against asteroid rotations faster than
about 2.2 h has been observed~\citep[e.g.,][]{prav07iaus}.

Although lightcurve observations provide powerful constraints on
binaries, there are limitations.  Detection of mutual events requires
an edge-on geometry and observations at the time of the events, such
that some binaries remain undetected (e.g., (69230) Hermes during its
2003 apparition).  Small satellites also escape detection because
their effect on the lightcurve is not measurable (e.g., satellites
with $D_s/D_p \lesssim$ 0.17 remain undetected if the minimum
detectable relative brightness attenuation is $\sim$0.03 mag).  The
probability of mutual event detection is larger at smaller semi-major
axes (expressed in units of primary radius) and at larger size ratios,
resulting in observational biases~\citep[e.g.,][]{prav12}.  Finally,
lightcurve observations yield relative, not absolute, measurements of
orbital separations.  Detection of small or distant secondaries and
direct measurement of orbital separation must instead rely on other
observational techniques.

\bigskip
\noindent
\textbf{ 2.3 Lightcurve Observations of Asteroid Pairs}\hspace{-1cm}
\bigskip

\citet{vokr08} reported evidence for pairs of MBAs with bodies in each
pair having nearly identical heliocentric orbits.  Because chance
associations can be ruled out, the asteroids in each pair must be
genetically related.  Quantifying the difference in orbital parameters
is accomplished with a metric $d$ that corresponds roughly to the
relative velocity between the bodies at close encounter.
\citet{vokr08} identified 44 asteroid pairs (excluding family members)
with a distance between the orbits of their components amounting to $d
< 10$~m/s.  They showed that, when integrated backwards in time, the
orbits converge 
at a certain moment in the past with a physical distance much less
than the radius of the Hill sphere and with a low relative velocity on
the order of 1~m/s.

\begin{figure}[ht!]
  \includegraphics[angle=0,width=\columnwidth]{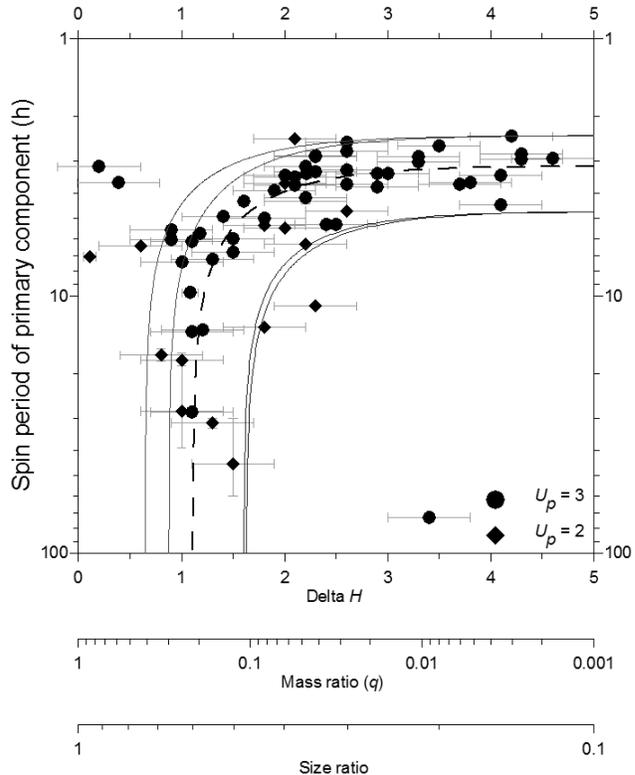}
  \caption{\small Primary rotation periods 
   versus mass ratios 
   of asteroid pairs.  The mass ratio values were estimated from the
   differences between the absolute magnitudes of the pair components,
   $\Delta H$.  Circles are data points with quality code rating
   $U_p=3$, meaning a precise period determination.  Diamonds are data
   points with $U_p=2$, which are somewhat less certain estimates.
   Error bars are one standard deviation.  The data match the
   predictions (curves) of a model of rotational fission with a few
   adjustable parameters.  In the model, $A_{\rm ini}$ is the binary
   system's initial orbit semi-major axis, $\alpha_L$ is the
   normalized total angular momentum of the system (Fig.~3), and
   $a_p$, $b_p$, $c_p$ are the long, intermediate, and short axis of
   the dynamically equivalent equal mass ellipsoid of the primary.
   All models shown assume 
   $b_p/c_p=1.2$.  The dashed curve shows the best-fit model with
   $\alpha_L = 1.0$, 
$a_p/b_p=1.4$ 
and 
   $A_{\rm ini}/b_p=3$.  Solid curves represent upper and lower
   limiting cases with $\alpha_L=0.7-1.2$.
Figure updated from~\citet{prav10}.
    \label{fig-pairs}}
\end{figure}

\citet{prav09} developed a method to identify probable asteroid pairs
by selecting candidate pairs with a similar distance criterion, then
computing the probability that each candidate pair emerged as a result
of a coincidence between two unrelated asteroids.  They identified 72
probable asteroid pairs, reproducing most of the 44 previously known
pairs.  Most of the new candidates were later confirmed to be real
pairs using backward integrations of their heliocentric orbits.

\citet{vokr08} proposed a few possible formation mechanisms for the
asteroid pairs: collisional disruption, rotational fission, and
splitting of unstable asteroid binaries.  \citet{prav10} conducted a
survey of the rotational properties of asteroid pairs, and they found
a strong correlation between the primary rotational periods and the
secondary-to-primary mass ratio (Fig.~\ref{fig-pairs}).  They showed
that this correlation fits precisely with the predictions of a model
by \citet{sche07fission} in which a parent body with zero tensile
strength undergoes rotational fission.  The model predicts that
primaries of low mass ratio pairs ($q \lesssim 0.05$) have not had
their spin substantially slowed down in the separation process and
should rotate rapidly with frequencies close to the fission spin rate.
The observed periods are between 2.4 and 5 h.  Primaries of medium
mass ratio pairs ($q = 0.05$ to $\sim 0.2$) have had their spin slowed
down according to the model because a substantial amount of angular
momentum was taken away by the escaped secondary.  This trend is
observed in the data (Fig.~\ref{fig-pairs}).  Finally, high mass ratio
pairs with $q > 0.2$ should not exist, as the free energy in the
proto-binary system formed by rotational fission would be negative and
the components would be unable to separate.  Observations mostly
corroborate this prediction: all 32 pairs in the sample of
\citet{prav10} were found to have a mass ratio $\lesssim$ 0.2.
However, an expanded photometric survey with 64 asteroid pairs
observed between 2012 and the date of this writing reveals 3 pairs
with high mass ratio ($q > 0.5$).  Their formation requires an
additional supply of angular momentum.  Another important finding by
\citet{prav10} is that the primaries of asteroid pairs have lightcurve
amplitudes that imply shapes with a broad range of elongations, i.e.,
unlike the primaries of binaries (Sections 2.1 and 2.2), the primaries
of asteroid pairs do not tend to be nearly spheroidal.

\bigskip
\noindent
\textbf{ 2.4 Spectral Observations of Asteroid Pairs}
\bigskip 

Colorimetric and spectral observations of about 20 asteroid pairs
indicate that members of an asteroid pair generally have similar
spectra~\citep{dudd12, mosk12, dudd13, poli14, wolt14}.  In some
pairs, the authors observed subtle spectral differences between the
components and attributed them to a larger amount of weathered
material on the surface of the primary.  In two pairs, they observed
somewhat more significant spectral differences.  For the pair
(17198)--(229056), both \citet{dudd13} and \citet{wolt14} found that
the primary is redder, i.e., it has a somewhat higher spectral slope
than the secondary in the observed spectral range 0.5--0.9~$\mu$m.  It
is unclear why their spectra differ despite a strong dynamical link
between the two asteroids.
For the pair (19289)--(278067), \citet{wolt14} observed a spectral
difference similar to that seen in (17198)--(229056), but
\citet{dudd13} observed very similar spectra.  Cross-validation of the
methods or additional observations, perhaps rotationally resolved, are
needed to resolve the discrepancy.

\begin{figure*}[!ht]
  \includegraphics[width=1\textwidth]{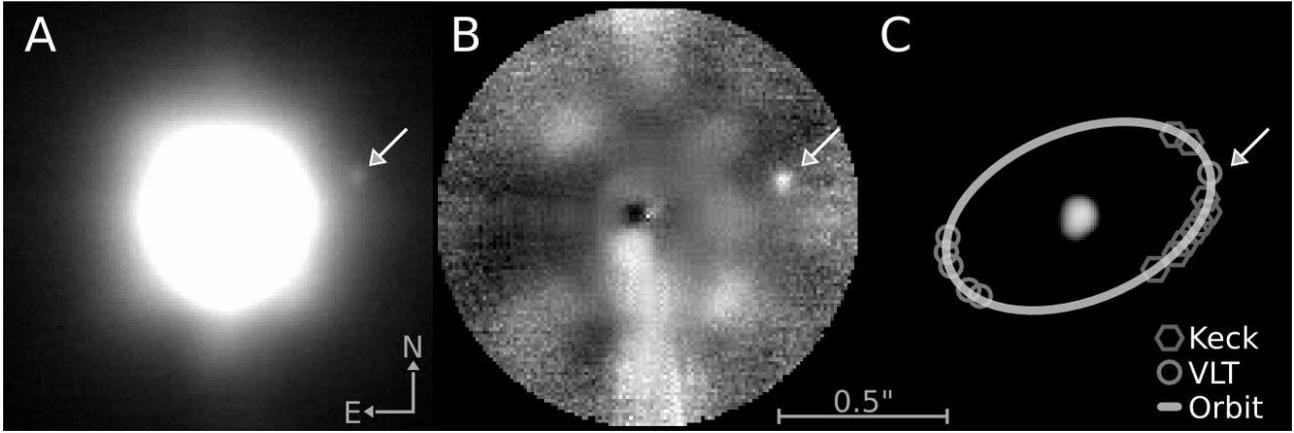}
  \caption{\small Satellite detection by direct imaging with adaptive optics (AO).
(a) Image of asteroid (41) Daphne (Vmag=10) obtained with a 
    ground-based AO-fed camera (NACO at ESO VLT, 5 s exposure).
(b) Same image after subtraction of the flux from the 
    primary, enabling more accurate measurements of the flux and 
    position of the secondary.  
(c) Orbit determination. The relative positions of
    the satellite from VLT/NACO and Keck/NIRC2 images are
    indicated.  Figure adapted from~\citet{carr09}.
    \label{fig-ao}}
\end{figure*}

\bigskip
\noindent
\textbf{ 2.5 Direct Imaging of MBA and Trojan Systems}
\bigskip
Direct imaging of asteroids can reveal the presence of satellites and,
following the long tradition of orbit determination of binary stars
and planetary satellites, lead to estimates of orbital parameters
(Fig.~\ref{fig-ao}).  This observing mode remains challenging because
the satellites are generally much smaller and fainter than their
respective primaries and because most satellites known to date orbit
at angular separations below 1 arcsecond.
Satellite discoveries have therefore followed the development of
adaptive optics (AO), and recent advances have enabled the detection
of asteroid satellites that had remained undetected in prior searches.

Instruments must have sufficient contrast and resolving power to
detect asteroid satellites with direct imaging.  For a 50--100~km
diameter asteroid in the main belt orbited by a satellite a few km
across, the typical angular separation is generally less than an
arcsecond with a contrast of 5 to 10 magnitudes (computed as 2.5 $\log
(F_p/F_s )$, where $F$ is the flux and $p$ and $s$ indicate primary
and secondary, respectively).

In some situations, direct images can actually resolve the primary.  A
50--100~km diameter asteroid at 2 au subtends 34--68 mas while the
diffraction limit of a 10 m telescope at a typical imaging wavelength
of 1.2 $\mu$m is about 30 mas.  Although the diffraction limit is not
reached, it can be approached with high-performance AO
instruments in excellent conditions.  With a sequence of disk-resolved
images that provide sufficient orientational coverage, it is possible
to estimate the 3D shape of the primary.
This enables volume and density determinations.
Instruments capable of meeting the contrast and resolution requirements 
include the Hubble Space Telescope (HST) and large (10~m class)
ground-based telescopes equipped with AO.
Spacecraft encounters provide an opportunity to detect small
satellites at small separations because of proximity to the target and
the absence of the point spread function halo that affects
ground-based AO instruments.

At the time {\em Asteroids III} was published, MBA satellite
discoveries included one by spacecraft ((243) Ida), one by HST ((107)
Camilla), and 6 by ground-based AO instruments.
Since then, ground-based AO instruments have been responsible for
almost all large MBA satellite discoveries:
(121) Hermione \citep{iauc7980}, 
(379) Huenna~\citep{iauc8182}, \
(130) Elektra \citep{iauc8183}, 
a second satellite to (87) Sylvia \citep{iauc8582} and to (45) Eugenia \citep{iauc8817}, 
(702) Alauda~\citep{cbet1016}, 
(41) Daphne \citep{iauc8930}, 
two satellites to (216) Kleopatra \citep{iauc8980} and (93) Minerva \citep{iauc9069}, 
and (317) Roxane \citep{iauc9099}.  
The wide binaries (1509) Esclangona~\citep{iauc8075} and (4674)
Pauling~\citep{iauc8297}, which are small asteroids in our
classification, have also been identified using AO-fed cameras.  HST
enabled detections of two additional wide binaries: (22899) 1999
TO$_{14}$~\citep{iauc8232} and (17246) 2000
GL$_{74}$~\citep{iauc8293}, both of which are small MBAs.  No
satellites have been discovered around any of the 7 asteroids recently
visited by spacecraft: (4) Vesta, (21) Lutetia, (2867) {\v S}teins,
(4179) Toutatis, (5535) Annefrank, (25143) Itokawa, and (132524) APL.
The number of known large MBAs with satellites is now 16, which
includes the only known large doubly synchronous system, (90)
Antiope~\citep{merl00,mich04,desc07,desc09}.  The fraction of large
MBAs with satellites is difficult to estimate because of a complex
dependence of satellite detectability on primary-to-secondary angular
separation and primary-to-secondary flux ratio.  However, because
several independent programs have surveyed over 300 large MBAs, it is
likely that the abundance of binaries in large MBAs is substantially
smaller than the $\sim$16\% abundance in NEAs and small MBAs.

\begin{figure}[p]
  \includegraphics[angle=0,width=0.97\columnwidth]{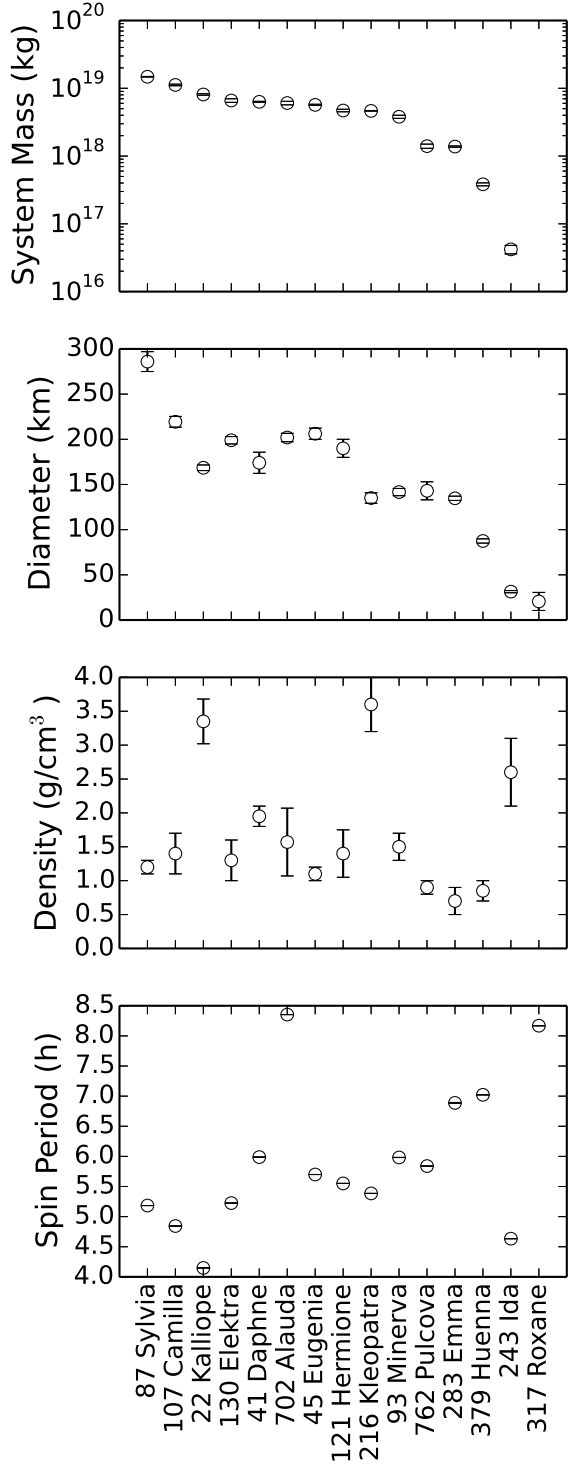}
  \caption{\small Properties of large MBA binaries and triples,
    excluding the doubly synchronous (90) Antiope.  Error bars or
    upper limits, when available, are shown.  Figures~\ref{fig-prim}
    and \ref{fig-sec} are based on data compiled by~\citet{john14} from references therein.
    \label{fig-prim}}
\end{figure}

\begin{figure}[p]
  \includegraphics[angle=0,width=\columnwidth]{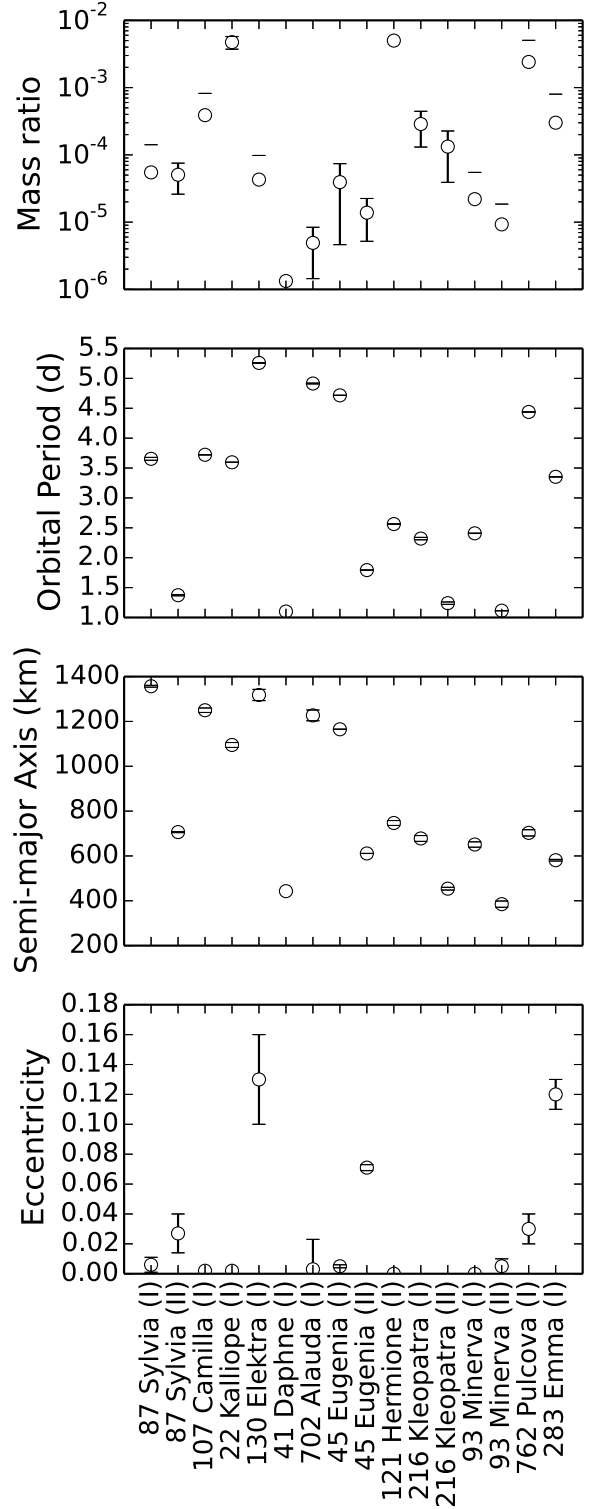}
  \caption{\small Properties of satellites of large MBAs, excluding
    outliers (90) Antiope and (379) Huenna (see text). Satellites of
    (243) Ida and (317) Roxane, whose orbits are not well known, are
    not shown.  %
    \label{fig-sec}}
\end{figure}

Properties of large MBA binaries and triples are summarized in
Figs.~\ref{fig-prim} and \ref{fig-sec}.
With the exception of the nearly equal-mass binary (90) Antiope, the
known satellites have secondary-to-primary mass ratios between
10$^{-6}$ and 10$^{-2}$.  All have orbital periods between 1 and 5.5
days, except (379) Huenna, whose orbit has a period of $\sim$88 days
and an eccentricity of $\sim$0.2~\citep{marc08}.  Many orbits have
near-zero eccentricity~\citep[e.g.,][]{marc08circular}, likely the
result of tidal damping, but the inner satellites of triples generally
have non-zero eccentricities.  These eccentricities may have
originated when orbits crossed mean motion resonances while tidally
expanding~\citep[e.g.,][]{fang12sylvia}.

At first glance, large MBA densities appear to cluster in two groups,
between 1 and 2~g/cm$^{3}$ and above 3~g/cm$^{3}$.  However,
interpretations are limited by the possibility of systematic errors,
including overestimates of volumes and underestimates of
densities~\citep{prav07}.  Because volume uncertainties almost always
dominate the error budget for binary asteroid
densities~\citep[e.g.,][]{merl02,carr12pss}, it is important to assess
the realism of uncertainties associated with volume determinations.
Some published density values should be regarded with caution because
overconfidence in the fractional uncertainty of volume estimates has
led to underestimates of bulk density uncertainties.  The platinum
standard of an orbiting spacecraft yields densities with $\sim$1\%
accuracy.
The gold standard of radar observations where tens of images with
hundreds or thousands of pixels per image are used to reconstruct a
detailed 3D shape model yields volumes (and densities) with $\sim$10\%
accuracy.  In contrast, AO images contain at most a few independent
resolution cells of the target asteroid.  Shape reconstructions based
on AO images and/or lightcurve data may not routinely yield volume
accuracies at the 10\% level, although one analysis reached that
level~\citep{carr12}.  In the absence of precise volume information,
one might be tempted to infer bulk densities from the theory of fluid
equilibrium shapes, but this approach is
problematic~\citep{hols07,harr09}.

In the Jupiter trojan population, one satellite to (624) Hektor has
been reported~\citep{iauc8732} since the discovery of the first trojan
satellite to (617) Patroclus~\citep{iauc7741}.  These are the only
trojans confirmed to have satellites in spite of several active search
programs.  The apparent low abundance of binary trojans is intriguing
and, if confirmed, may provide additional support for the idea that
Jupiter trojans originated in the trans-Neptunian
region~\citep{morb05,levi09} where they experienced a different
collisional environment than in the main belt of asteroids.
(624) Hektor has a satellite in a
$\sim$3-day orbit that is eccentric ($\sim$0.3) and inclined
($\sim$50$^\circ$) with respect to Hektor's equator~\citep{marc14}.
(617) Patroclus is unusual because it has two components of similar
size in a relatively tight ($\sim$680~km) orbit, with a normalized
total angular momentum exceeding that available from fission of a
single parent body~\citep{marc06}.
In the trans-Neptunian region, 14 and 64 binary systems have been
discovered with AO and HST, respectively~\citep{john14}.  The apparent
larger abundance of binary TNOs in the cold classical belt may be due
to a different dynamical environment and formation mechanism (Section 5).

Objects in the trojan and TNO populations are generally too faint for
AO observations in natural guide star (NGS) mode, in which the science
target is also used to measure the properties of the wavefront and 
command the deformable mirror.  These objects can be observed in
appulse when their sky position happens to be within $\lesssim$ 1
arcminute of a bright star.
The advent of laser guide star (LGS) adaptive optics has been an
important development that has freed the observer from finding such
chance alignments and has opened up a larger fraction of the sky for
observation of faint objects.  Even with LGS, however, the
availability of a tip-tilt star (Rmag $\lesssim$ 18) within $\lesssim$
1 arcminute of the target is still required.

\enlargethispage{0.5cm}

High-resolution and high-contrast imaging capabilities are
aggressively sought by instrument builders, in part to enable direct
imaging of exoplanets.  Cameras equipped with high-performance AO are
currently being installed or commissioned on large ground-based
telescopes: HiCIAO on Subaru, GPI on Gemini, and SPHERE at the ESO
VLT.  These instruments will improve the ability to detect faint
satellites orbiting close to their respective primaries.  However, in
most cases, asteroids fall in the faint-end range of these instrument
capabilities.  The next generation of large telescopes ($\sim$30~m
diameter) such as the Thirty Meter Telescope (TMT) and European
Extremely Large Telescope (E-ELT) will provide an improvement in
sensitivity by a factor of $\sim$10 and in angular resolution by a
factor of $\sim$3 compared to current 10 m telescopes.
With the anticipated development of AO capabilities at shorter
wavelengths, the second generation of instruments at these facilities
is expected to provide improvements in angular resolution by a factor
of $\sim$5.  Such instruments may allow detection of the small MBA
binaries that are currently beyond the reach of direct imaging
instruments.  In many of these systems, the components are separated
by only a few mas and the size ratios are larger than in large MBA
binaries, resulting in flux ratios closer to unity.

\pagebreak
\bigskip
\noindent
\textbf{ 2.6 Spectral Observations of MBA and Trojan Systems}
\bigskip

It is generally difficult to separate the light emitted or reflected
from the secondary from that of the brighter primary.  Nevertheless,
such observations can be attempted when the secondary happens to be at
a large angular separation from the primary, when the system is
undergoing mutual events, or with the help of an integral field
spectrograph.

Spectra of (22) Kalliope and its satellite Linus in the 1--2.4 $\mu$m
region appear to be similar~\citep{lave09}, which the authors
attribute to satellite formation after a major impact on the precursor
body.  Observations of both components of (90) Antiope in the same
spectral region also shows surface reflectances that are
similar~\citep{marc11}.  The spectrum of (379) Huenna is
characteristic of C-type asteroids and the secondary does not exhibit
a significantly different taxonomic type~\citep{deme11}.  Both
components of (809) Lundia are consistent with a V-type
classification~\citep{birl14}.

In the mid-infrared, Spitzer observations of the trojan (617)
Patroclus, including during mutual events, provided size estimates for
its components and a thermal inertia of 20 $\pm$ 15
J~s$^{-1/2}$~K$^{-1}$~m$^{-2}$ \citep{muel10}.  Spitzer observations
combined with photometric results in the visible yielded size and
albedo estimates for (624) Hektor ~\citep{emer06}.  %
Spitzer observations of these and other binaries did not resolve the
binaries and results typically cannot be compared to observations that
place many resolution elements on individual components.  One
exception is 2000 DP$_{107}$, where analysis of Spitzer data yields a
system density of $0.9 \pm 0.3$~g/cm$^{3}$~\citep{marc12} and the
radar results indicate $1.4 \pm 0.2$~g/cm$^{3}$~\citep{naid15dp}.

\bigskip
\noindent
\textbf{ 2.7 Stellar Occultations of MBA and Trojan Systems}
\bigskip

\begin{figure}[!ht]
  \centering
  \includegraphics[width=.45\textwidth]{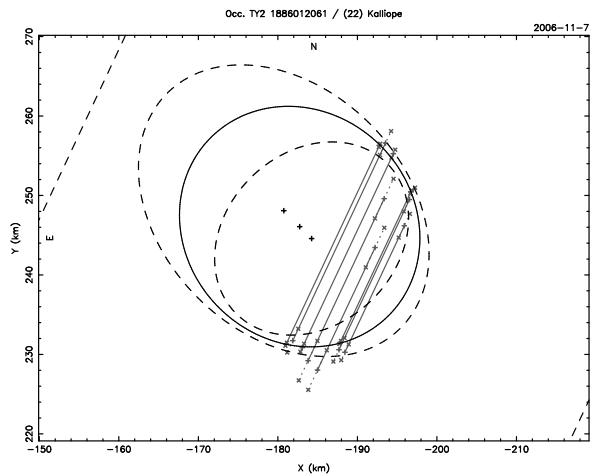}
  \caption{\small 
The apparent shape of Linus, a satellite of (22)
Kalliope~\citep{marg03s}, detected by stellar occultations.  In this
analysis, the profile of the satellite (solid curve) fitted to the
observed chords (straight lines) yields an equivalent diameter of 30
$\pm$ 6 km.  Dashed curves show the corresponding uncertainty of the
fitted profile, and dashed lines show negative detections.  Figure
adapted from \citet{2008-Icarus-196-Descamps}.
    \label{fig-occ}}
\end{figure}

Stellar occultations provide a way of detecting components of a
multiple-asteroid system, of placing bounds on component sizes, and of
obtaining the relative positions of components on the plane of the
sky.  A recording of star light as a function of time shows a deep
extinction when a target body crosses the line of sight between the
observer and the star.  This can be interpreted in terms of a {\em
  chord} on the apparent disk of the target body projected on the
plane of the sky.  If several observers are placed across the
occultation path on the surface of the Earth, multiple chords can be
obtained, and the size and shape of the target projected on the sky
can be reconstructed (Fig.~\ref{fig-occ}).  When two or more
components are present, it is also possible to measure their relative
position.  While the reliability of this technique was disputed a
decade ago due to the lack of digital recordings, the availability of
low-cost cameras and global positioning systems has enabled a dramatic
improvement in the precision of timing reports.  Stellar occultations
have become an important observational tool for the study of binary
asteroids.

Early reports~\citep[e.g.,][]{binz79} of asteroids suspected to be
binaries on the basis of occultation data (including (3) Juno, (6)
Hebe, (9) Metis, (12) Victoria, (18) Melpomene, (146) Lucina, and
(532) Herculina) have remained largely unconfirmed despite extensive
follow-up searches.  However, it is likely that the outer satellite of
(216) Kleopatra was detected during a 1980
occultation~\citep{dunh81,2011-Icarus-211-Descamps}.  The detection of
a satellite around the trojan (911) Agamemnon has been
suggested~\citep{2013-PSS-87-Timerson} but not yet confirmed.  The
occultation technique has also been used to detect rings around the
centaur (10199) Chariklo \citep{2014-Nature-508-Braga-Ribas}.

One strength of the stellar occultation technique lies in the fact
that the observability of the event depends mainly on the brightness
of the star and not of the asteroid or satellite. Stellar occultations
can thus be used to detect small (km size) satellites, even those that
are close to the primary and that would remain undetected in direct
imaging.

Another strength of the technique is the potential for high-precision
measurements.  Stellar occultations are based on time-series
photometry.  Given a sufficiently high cadence (e.g., 10--30 images per
second), it is possible to obtain a precision of a few mas on the
relative position of binary components, which is 5 to 10 times better
than with direct imaging with current instrumentation.

Finally, well-sampled stellar occultations allow for recovery of the
size and apparent shape of asteroids and their satellites, whereas
optical lightcurves and direct imaging observations provide primarily
the diameter ratio of the components and more limited shape
information.  So far, four successful observations of satellite size
and shape have been reported: Linus, satellite of (22) Kalliope
\citep{2008-Icarus-196-Descamps}, Romulus, the outer satellite of (87)
Sylvia \citep{2014-Icarus-239-Berthier}, and both components of the
equal-size binaries (90) Antiope~\citep{2014-MNRAS-443-Bartczak} and
(617) Patroclus~\citep{buie14}.

Despite all of these strengths, there remains a relatively low number
of well-covered stellar occultation events. This is due, in part, to
the requirement of successful observations at many stations.  Owing to
uncertainties on both the star and asteroid positions, the occultation
path can shift by several tens or even hundreds of km on Earth
compared to the prediction. Observers must therefore spread
geographically to cover an event, but the detection of a satellite by
several stations requires a fine grid of observers.

The situation is, however, expected to improve dramatically with the
availability of the Gaia stellar catalog and better asteroid orbits
\citep{2007-AA-474-Tanga}. Predictions of the occultation paths (for
the center of mass) will be accurate to a few km,
and the main source of uncertainty will become the prediction of the
relative position of the satellite around the primary.

\bigskip
\noindent
\textbf{ 2.8 Other Observations}
\bigskip

There have been several attempts to use ground-based interferometers
to measure the angular separation of binary systems
\citep{2009-ApJ-694-Delbo, carr15}.  However, asteroid
satellites are too faint for current interferometers operating in the
visible and near-infrared
and at the edge of detection in the mid-infrared.
Future instrumentation 
may allow such observations.  There are
also prospects for observations with the ALMA sub-millimeter array
\citep{2009-Icarus-200-Busch}.

\bigskip

\centerline{\textbf{ 3. DYNAMICS}}
\bigskip

In parallel with advances in instrumentation and observing
capabilities, the field has seen tremendous developments in
understanding the dynamical processes that affect asteroid systems.
This has been enabled in large part by the availability of detailed
shape models and orbital parameters, by the need to model the dynamics
of newly discovered triple systems, and by the desire to understand
formation and evolution processes.

A non-exhaustive list of some dynamical problems that have been
explored since {\em Asteroids III} includes the stability of asteroid
satellite orbits \citep{sche02b,2012-Icarus-220-Frouard}, the dynamics
around triaxial bodies \citep{sche09}, the fate of asteroid ejecta
\citep{sche07fission}, the formation of contact binaries via dynamical
evolution \citep{sche09,tayl11,tayl14},
the genesis of eccentric and mutually inclined orbits
\citep{fang11,fang12spinorbit}, the orbital determination of triple
systems using point-mass approximations
\citep{2010-Icarus-210-Marchis} and full N-body calculations
\citep{fang12sylvia}, the influence of Kozai cycles on
binaries~\citep{pere09,fang12kozai}, the effects of close planetary
encounters on mutual orbits \citep{fang12encounters} and spin states
\citep{taka13}, the complex spin-orbit interactions with irregular
component shapes \citep{sche06}, including the libration and irregular
rotation of secondaries \citep{naid15}, the influence of internal
structure~\citep{gold09}, material properties~\citep{tayl11} and
nonspherical shapes~\citep{tayl14} on tidal evolution, the possibility
of tidal saltation \citep{harr09,fahn09}, the possibility of
significant radiative evolution
\citep{cuk05,cuk07,cuk10,mcma10cmda,mcma10icarus}, and the possibility
of a stable equilibrium between tidal and radiative evolution
\citep{jaco11apj}.  

\begin{figure}[ht!]
\includegraphics[angle=0,bb=78 360 558 720,width=1\columnwidth]{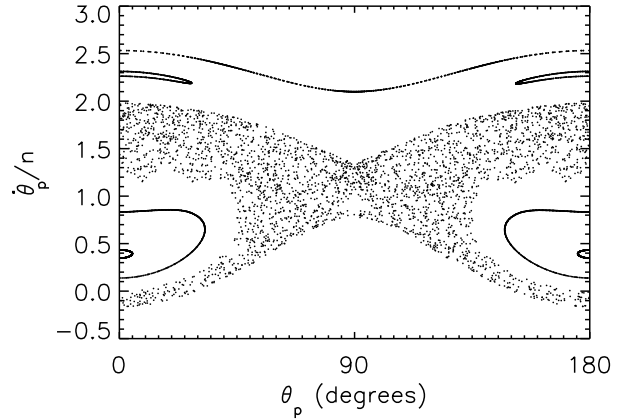}
\caption{\small 
Surface of section plot showing the possible rotational
regimes of the $\sim$200 m secondary of 1991 VH (secondary elongation
$a/b=1.5$ and mutual orbit eccentricity $e=0.05$).  The plot shows the
angle between the long axis and the line of apsides of the mutual
orbit, $\theta_p$, against its time derivative, $\dot{\theta_p}$,
normalized by the mean motion, $n$, at each pericenter passage.  
Five trajectories are illustrated (from top to bottom: non-resonant
quasi-periodic, periodic, chaotic, periodic, periodic).  While trapped
in the sea of chaos, the secondary experiences torques on its
permanent deformation that result in a highly variable spin rate,
preventing BYORP-type evolution.  Figure from \citet{naid15}.}
\label{fig-chaos}
\end{figure}

Several radar data sets provide exquisite constraints for dynamical
studies.  Reflex motion has been measured for 2000
DP$_{107}$~\citep{marg02s,naid15dp}, 
1999 KW$_{4}$~\citep{ostr06}, and 1991 VH~\citep{naid12dda}, allowing
masses of individual components to be determined.  Because detailed
component shapes are also available, one can fully model the system
dynamics and study spin-orbit coupling in
detail~\citep{sche06,fahn08,naid15}.  One finding from this work is
that even moderately elongated secondaries on mildly eccentric orbits
are likely to experience chaotic rotation that substantially affect
binary evolution timescales (Fig.~\ref{fig-chaos}).

\bigskip

\centerline{\textbf{ 4. SMALL ASTEROIDS: SYNTHESIS}}
\bigskip

\noindent
{\textbf{ 4.1. Rotational Fission Hypothesis}}
\bigskip

With the exception of the doubly synchronous binary asteroid systems,
the primary asteroids of all small binary systems are rapidly rotating
(within a factor of only a few of the critical disruption spin limit
for bodies with no shear or tensile strength $\omega_d = \sqrt{4 \pi \rho G / 3}$). 
Furthermore, almost all known small binary asteroids have
high angular momentum contents~\citep{prav07}.  
These characteristics are not consistent with formation following a
sub-catastrophic impact, capture through a three-body interaction in the
near-Earth or main belt, or capture after a catastrophic impact.
Instead, they are indicative of formation from a rotational fission
event~\citep[e.g.,][]{marg02s,prav07}.  The rotational fission
hypothesis posits that a parent asteroid can be torqued to a rotation
rate so great that the centrifugal accelerations overpower the
gravitational accelerations holding a
strengthless asteroid together~\citep{weid80}. It is possible that
some 
small asteroids have cohesive or molecular strength in addition to
self-gravity~\citep[e.g.,][]{rozi14}.  In these cases, the centrifugal
accelerations must overcome these additional forces in order for the
asteroid to fission~\citep{prav00fastslow,Sanchez:2014ir}.
At rapid rotation rates, loose surface material can flow from
high-latitude regions to the equator along potential
gradients~\citep{ostr06}.  It has been shown that rotational
acceleration could trigger local slope failures and landslides, which
can form the canonical top shape and equatorial bulge seen on primary
components in small multiple-asteroid systems~\citep{wals08n,harr09}.

~\citet{bott02} proposed a YORP-induced rotational fission
hypothesis. It has since been shown that the YORP effect controls the
rotational acceleration of small
asteroids~\citep{Bottke:2006en,Marzari:2011dx} and naturally explains
the period distribution among small
asteroids~\citep{prav08,Rossi:2009kz,poli09}. Furthermore, including
the YORP-induced rotational fission hypothesis in size-frequency
distribution models improves the agreement with
observations~\citep{jaco14dist}.  The observed characteristics of
the systems described in Sections 2.1-2.3 as well as thermal inertia
observations~\citep{delb11} are consistent with a binary formation
mechanism that involves spin-up and mass shedding.  The YORP-induced
rotational fission hypothesis is the leading candidate for explaining
the formation of binaries, triples, and pairs among small asteroids.

\bigskip
\noindent
{\textbf{ 4.2. Asteroid Pairs}}
\bigskip

The YORP effect can increase the spin rate of asteroids beyond the
critical disruption spin limit, thereby triggering rotational fission.
In actuality, there is some uncertainty regarding the spin rate at
which disruption occurs---there may be failure and deformation before
fission~\citep{wals08n,Sanchez:2011kw,CottoFigueroa:2013vu}.  The
critical disruption spin limit also depends on the detailed shapes,
masses, interlocking nature of the
interior components and any cohesive
forces~\citep{sche07fission,sche09min,Sanchez:2014ir}. Despite
ignoring these details, simple calculations provide a rotational
fission model that can be compared directly and successfully with
observations.

If a spherical approximation of each component is made, then the
rotational breakup spin rate $\omega_q$ necessary for fission as a
function of the secondary-to-primary mass ratio
$q$ is~\citep{sche07fission}:
\begin{equation}
\omega_q = \omega_d \sqrt{\frac{1+q}{\left(1 + q^{1/3}\right)^3}}.
\end{equation} 
This is the exact solution for two spheres resting on each other with
a mass ratio of $q$ and rotating about the axis of maximum moment of
inertia.

The spherical component model described above demonstrates the
important reality that the larger the mass ratio $q$ of the two future
binary members, the slower the required rotation rate necessary to
create the binary system. This slower required rotation rate
translates into a small initial free energy for the ensuing binary
system. The free energy $E_f$ is the energy that is accessible to the
different energy reservoirs in the system, including the rotation
states of each member and the orbit.  It does not include the internal
binding energy of each object. The free energy is an important
quantity because it determines the boundedness of the system. Bound
systems have negative free energy, while unbound systems have positive
free energy. An unbound binary system implies that the system is
capable of disruption but does not imply that the system will
disrupt. For the idealized case of two spheres, the free energy can be
expressed as~\citep{sche07fission}:
\begin{equation}
E_f = \frac{ 2 \pi \rho \omega_d^2 R_p^5}{15 } f(q),
\end{equation}
\noindent where $R_p$ is the radius of the primary and $f(q)$ is an
algebraic, monotonically decreasing function 
for $0<q\leq1$. For the equation above corresponding to two spheres,
the function crosses zero when
$q \approx 0.204$.  Similar equations can be written for any two
component shapes, but $q \sim 0.2$ remains near the binding energy
transition point, so the model uses this point as a simple
approximation. This crossing point divides bound systems with negative
energy and mass ratios $q > 0.2$ and unbound systems with positive
energy and mass ratios $q < 0.2$. Because of this fundamental
difference, high mass ratio $q > 0.2$ and low mass ratio $q < 0.2$
binary systems evolve
differently~\citep{sche09,jaco11icarus}. Primarily,
positive energy low mass ratio systems will chaotically explore
orbital phase space until the majority find a disruption trajectory
creating an asteroid pair; this evolutionary route is unavailable to
high mass ratio systems.

The asteroid pair population provides a natural laboratory to test
this relationship~\citep{sche07fission,vokr08}.  ~\citet{prav10}
examined many asteroid pair systems and measured the rotation rate of
the primary and the absolute magnitude difference between the pair
members. These two quantities should follow a simple relationship
related to $\omega_q$, although many of the ignored details mentioned
at the beginning of this section can move asteroids away from this
relationship. Indeed,~\citet{prav10} discovered that asteroid pairs do
follow this relationship (Fig.~\ref{fig-pairs}). Furthermore, they
found that the large members of asteroid pairs 
have a broader range of elongations than the primaries of binary
systems, consistent with the findings of~\citet{jaco11icarus} that
prolate primaries are less likely to remain in a bound binary system
after rotational fission.
Thus, there is strong evidence to support the hypothesis that asteroid
pairs are the products of rotational fission.

Asteroid pairs continue to be a fertile observational landscape. Since
dynamical integrations can derive the ``birthdate'' of such systems,
observers can test ideas regarding space weathering timescales and
YORP evolution after fission~\citep{poli14,Polishook:2014bb}. Along
with binary systems, the surfaces of asteroid pairs may provide clues
in the future regarding the violence of the rotational fission
process~\citep{poli14dust}.

\bigskip
\noindent
{\textbf{ 4.3. Binary and Triple Systems}}
\bigskip

\begin{figure*}[!htb]
\centering
\includegraphics[width=\textwidth]{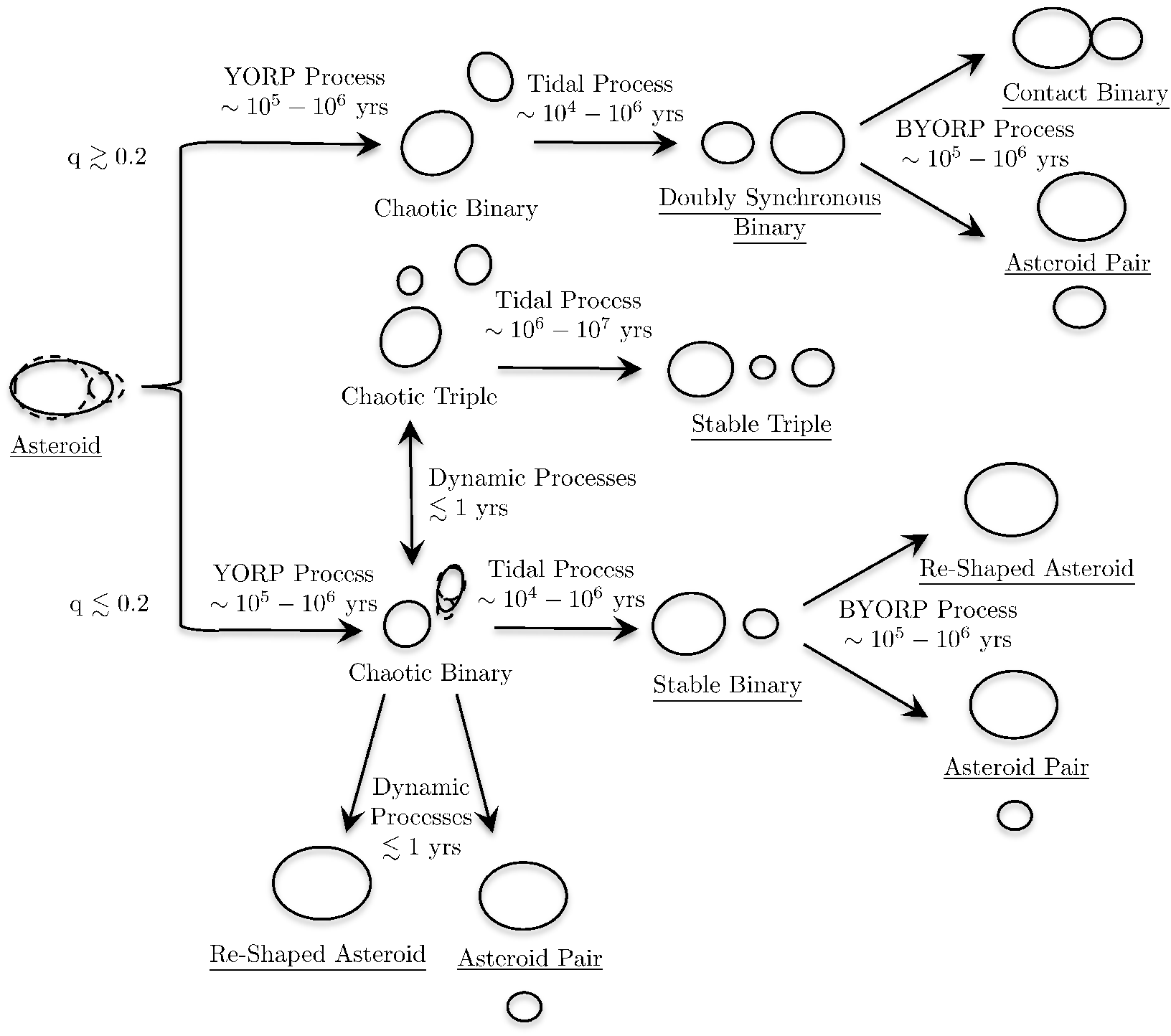}
\caption{\small 
Flowchart showing the possible evolutionary paths for an
asteroid after it undergoes rotational fission. Each arrow is labeled
with the dominant process and an estimated timescale for this
process. Underlined states are nominally stable for a YORP effect
timescale. Figure from \citet{jaco11icarus}.}
\label{fig-AsteroidFlowChart}
\end{figure*}

\citet{jaco11icarus} showed that after rotational fission there are
a number of possible
outcomes. Their numerical studies produced the evolutionary flow chart
shown in Fig.~\ref{fig-AsteroidFlowChart}; many of these outcomes
were also found by~\citet{fang12spinorbit}. The high and low mass
ratio distinction for rotational fission emphasized above plays an
important role in distinguishing the two evolutionary pathways. Along
the high mass ratio pathway, both binary members tidally synchronize
and then evolve according to the BYORP effect.

Along the low mass ratio pathway, the binary system is unbound. Since
these systems are chaotic, many are disrupted and become asteroid
pairs. During this chaotic binary state, the secondary can often go
through rotational fission itself, although this rotational fission is
torqued by spin-orbit coupling (Fig.~\ref{fig-chaos}) rather than the
YORP effect.
Loss of material from the secondary stabilizes the remaining orbiting
components.  The lost mass may reaccrete onto the primary, perhaps
contributing to the observed equatorial ridges, or may escape from the
system.  
In these cases, the system undergoes another chaotic binary episode
with three possible outcomes: a re-shaped asteroid, an asteroid pair, or
a stable binary.
These binaries still possess positive free energy such that they may
disrupt if disturbed.  In other cases, the system retains three
components after secondary fission.  While the numerical simulations
of \citet{jaco11icarus} did not yield this latter outcome, it is possible
that this pathway explains the existence of stable triple systems.
After stabilization of the low mass ratio binary system, the secondary
synchronizes due to tides~\citep[e.g.,][]{gold09},
although some satellites may be trapped in a chaotic rotation state
for durations that exceed the classic spin synchronization
timescales~\citep{naid15}.  Then the system evolves according to the
BYORP effect and tides. These binary evolutionary processes and their
outcomes are discussed in Walsh \& Jacobson (this volume).  As shown
in Fig.~\ref{fig-AsteroidFlowChart}, these evolutionary paths include
each of the binary morphologies identified in this chapter and by
other teams~\citep{prav07,fang12spinorbit}.  In particular, the
formation of wide asynchronous binaries such as (1509) Esclangona,
(4674) Pauling, (17246) 2000 GL$_{74}$, and (22899) 1999 TO$_{14}$ is
best explained by a rotational fission mechanism~\citep{poli11}
followed by BYORP orbital expansion~\citep{jaco14wide}.

An alternative
formation mechanism for triples such as (153591) 2001 SN$_{263}$ and
(136617) 1994 CC is that after creating a stable binary system, the
primary undergoes rotational fission a second time. As long as the
third component is on a distant enough orbit, then this process may
result in a stable triple
system~\citep{fang11,fang12spinorbit,jaco14wide}.

\bigskip

\begin{figure}[!htb]
  \includegraphics[angle=-90,width=\columnwidth]{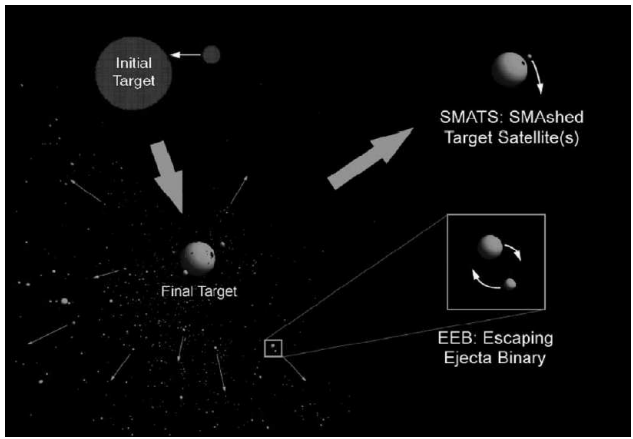}
  \caption{\small Numerical simulations show that binaries can form as
    a result of large impacts between asteroids.  In some scenarios,
    impact debris can remain gravitationally bound to the target body,
    forming a satellite (SMATs).  This process likely explains the
    formation of large MBA binaries.  In other scenarios, two
    fragments from the escaping ejecta have sufficiently similar
    trajectories, such that they become bound to one another (EEBs).
    Figure from~\citet{durd04}.
    \label{fig-durda}}
\end{figure}

\centerline{\textbf{ 5. LARGE ASTEROIDS: SYNTHESIS}}
\bigskip
The primaries of most known binary and triple asteroids greater than
20~km have spin periods in the range of 4 h to 7 h
(Fig.~\ref{fig-prim}).  While these spin rates are not near the
disruption spin limit, they are typically faster than the mean spin
rates for asteroids of similar sizes.  The total angular momentum
content, however, is well below that required for rotational fission.
The secondary-to-primary mass ratios in these systems range from
10$^{-6}$ to 10$^{-2}$.  These properties are consistent with
satellite formation during large collisions (Fig.~\ref{fig-durda}).
\citet{durd04} have shown in numerical simulations that impacts of 10-
to 30-km diameter projectiles striking at impact velocities between 3
kms$^{-1}$ and 7 kms$^{-1}$ can produce satellites that match observed
properties.  Multiple asteroid systems, e.g., (45)
Eugenia~\citep{merl99,iauc8817} and (87)
Sylvia~\citep{marg01dps,marc05} can also plausibly form through
collisions.

There is more uncertainty related to the formation of (90) Antiope and
(617) Patroclus, which are both too large to be substantially affected
by YORP.  Hypotheses for the formation of (90) Antiope include
primordial fission due to excessive angular momentum~\citep{prav07},
an improbable low-velocity collision of a large
impactor~\citep{weid01}, or shrinking of an initially wide binary
formed by gravitational collapse~\citep{nesv10}.  Gravitational
collapse in a gas-rich protoplanetary disk has been invoked to explain
the formation of numerous binaries in the trans-Neptunian region.
(617) Patroclus may be a primordial TNO that avoided disruption during
emplacement in the trojan region~\citep{nesv10}.  Wide TNO binaries
would not be expected to survive this process, whereas encounter
calculations~\citep[e.g.,][]{fang12encounters} show that tight
binaries would.

\bigskip

\centerline{\textbf{ 6. CONCLUSIONS}}
\bigskip
Studies of binaries, triples, and pairs remain a fertile ground for
observing processes that are important in planet formation and for
measuring quantities that are difficult to obtain by other means.
These include masses and densities as well as thermal, mechanical, and
interior properties.  Binaries or triples have been found in $\sim$50
NEAs, $\sim$50 small MBAs, $\sim$20 large MBAs, and 2 trojans.  A
unifying paradigm based on rotational fission and post-fission
dynamics explains the formation of small binaries, triples, and pairs.
Because the sun-powered rotational fission process is unrelenting, and
because the production of pairs is a frequent outcome of this process,
a substantial fraction of small bodies likely originated in a
rotational disruption event.  This origin affects the size
distribution of asteroids and may explain the presence of single NEAs
with equatorial bulges observed with radar.  Small satellites of large
MBAs are likely formed during large collisions.  Advances in
instrumentation, observational programs, and analysis techniques hold
the promise of exciting findings in the next decade.

\bigskip

\bibliographystyle{plainnat}

\bibliography{bin}

\end{document}